\documentclass[a4paper,12pt]{article}
\usepackage{amsmath}
\usepackage{amssymb}
\usepackage{,amsfonts,amsthm,dsfont,dcolumn}
\usepackage{caption}
\usepackage{graphicx,wrapfig}
\usepackage{subfigure}
\usepackage[hypertex]{hyperref}

\DeclareGraphicsExtensions{.pdf,.png,.jpg,.jpeg,.eps}
\newcommand{\beq}{\begin{equation}}
\newcommand{\eeq}{\end{equation}}
\newcommand{\dee}{\partial}

\newcommand{\al}{\alpha}

\newcommand{\ga}{\gamma}

\newcommand{\mo}{{\mathcal {O}}}

\newcommand{\beg}{\begin{gathered}}
\newcommand{\eeg}{\end{gathered}}
\newcommand{\bal}{\begin{align}}
\newcommand{\eal}{\end{align}}
\newcommand{\bea}{\begin{eqnarray}}
\newcommand{\eea}{\end{eqnarray}}
\newcommand{\nn}{{\nonumber}}

\begin{document}
\begin{titlepage}
\thispagestyle{empty}
\begin{flushright}
UK/10-11
\end{flushright}

\bigskip

\begin{center}
\noindent{\Large \textbf
{Integrability Lost:}}\\
{\bf Chaotic dynamics of classical strings on a confining holographic background}\\
\vspace{2cm} \noindent{
Pallab Basu\footnote{e-mail:pallab.basu@uky.edu},
Diptarka Das\footnote{e-mail:diptarka.das@uky.edu} and
Archisman Ghosh\footnote{e-mail:archisman.ghosh@uky.edu}}

\vspace{1cm}
  {\it
 Department of Physics and Astronomy, \\
 University of Kentucky, Lexington, KY 40506, USA\\
 }
\end{center}

\vspace{0.3cm}
\begin{abstract}

It is known that classical string dynamics on pure $AdS_5\times S^5$ is integrable and plays an important role in solvability. This is a deep and central issue in holography. Here we investigate similar classical integrability for a more realistic confining background and provide a negative answer. The dynamics of a class of simple string configurations on $AdS$ soliton background can be mapped to the dynamics of a set of non-linearly coupled oscillators. In a suitable limit of small fluctuations we discuss a quasi-periodic analytic solution of the system. Numerics indicates chaotic behavior as the fluctuations are not small. Integrability implies the existence of a regular foliation of the phase space by invariant manifolds. Our numerics shows how this nice foliation structure is eventually lost due to chaotic motion. We also verify a positive Lyapunov index for chaotic orbits. Our dynamics is roughly similar to other known nonintegrable coupled oscillator systems like H\'enon-Heiles equations.

\end{abstract}
\end{titlepage}
\newpage

\newpage


\section{Introduction} 

$AdS$/CFT duality \cite{Maldacena:1997re} was a major step towards the goal of recasting large-$N$ QCD as a string theory. One of the interesting aspects of $AdS$/CFT is integrability (see the review \cite{Beisert:2010jr}).  Integrability has allowed us to obtain many classical solutions of the theory that would otherwise have been impossible to find \cite{Gubser:2002tv,Tseytlin:2010jv}. One may approach integrability from two sides corresponding to the two extreme values of the 't Hooft coupling. On the supergravity side (which is a good description as $\lambda \to \infty$), integrability of the classical sigma model on $AdS_5\times S^5$ was established for bosonic sector in \cite{Mandal:2002fs} and fully completed with the inclusion of fermions in \cite{Bena:2003wd}. It has been shown that classical string motion in $AdS_5\times S^5$ has an infinite number of conserved charges.\footnote{However the commutator algebra of such charges is not fully understood.} This is the closest to solvability that we can get currently. It is also to be noted that the study of integrability in non-linear sigma models has a long history \cite{Luscher:1977rq,Brezin:1979am}. The other approach to integrability is perturbative in the weakly coupled gauge theory \cite{Beisert:2010jr}. We will restrict ourselves to the former approach. 

An important open question is whether integrability can be extended to more QCD-like theories.\footnote{This is one of the motivations discussed in the introduction of \cite{Bena:2003wd}.} The original form of $AdS$/CFT duality was for conformally invariant ${\cal N}=4$ SYM theory but this can be deformed in various ways to produce string duals to confining gauge theories with less or no supersymmetry. The construction of \cite{Mandal:2002fs, Bena:2003wd} does not readily generalize to these less symmetric backgrounds. One prime example of a confining background is the $AdS$ soliton \cite{Horowitz:1998ha,Witten:1998zw}. Similar geometries have been used extensively to model various aspects of QCD in the context of holography \cite{Sakai:2004cn}. Here we will look at the question of integrability of bosonic strings on an $AdS$ soliton background. Although it is much more interesting to explore full quantum integrability, to begin with we may ask whether we can find enough conserved charges even at a purely classical level. The answer turns out to be negative. By choosing a class of simple classical string configurations, we show that the Lagrangian reduces to a set of coupled harmonic and anharmonic oscillators that correspond to the size fluctuation and the center of mass of motion of the string. The oscillators decouple in the low energy limit. With increasing energy the oscillators become nonlinearly coupled. Many such systems are well known to be chaotic and nonintegrable \cite{Ott,Hilborn}. It is no surprise that our system also shows a similar behaviour. Possibly chaotic behaviour of a test string has been argued previously in black hole backgrounds \cite{Zayas:2010fs,Frolov:1999pj}. However our problem is somewhat different as we are looking at a zero temperature geometry without a horizon. In a companion paper \cite{Basu:2011} non-integrability of string theory in $AdS_5 \times T^{1,1}$ is discussed. 
 
We start by discussing the $AdS$ soliton background and our test string ansatz$^{\S\ref{sec:setup}}$. We discuss some quasi-periodic solutions for small oscillation regime. Our argument for the non-integrability is via numerical solution of the EOM's$^{\S\ref{sec:dyn}}$. In a certain regime of parameter space the system shows a zigzag aperiodic motion characteristic of a chaotic system. We then look at the phase space. Integrability implies the existence of a regular foliation of the phase space by invariant manifolds, known as KAM (Kolmogorov-Arnold-Moser) tori, such that the Hamiltonian vector fields associated with the invariants of the foliation span the tangent distribution. Our numerics shows how this nice foliation structure is gradually lost as we increase the energy of the system$^{\S\ref{subsec:poin}}$. To be complete we also calculate Lyapunov indices for various parameter ranges and find large positive values in chaotic regimes$^{\S\ref{subsec:lya}}$. We discuss open questions and possible extensions in the conclusion$^{\S\ref{sec:con}}$.

\section{Setup}\label{sec:setup}
The $AdS$ soliton ($\cal M$) metric for an asymptotically $AdS_{d+2}$ background is given by \cite{Horowitz:1998ha},
\begin{align}
& ds^2=L^2 \alpha' \left(e^{2u}(-dt^2+T_{2\pi}(u) d\theta^2+dw_i^2)+\frac{1}{T_{2\pi}(u)} du^2 \right)\,, \nn \\
{\rm where\ } & \ T_{2\pi}(u)=1-{\left(\frac{d+1}{2} e^u\right)}^{-(d+1)}\,.
\label{metric}
\end{align}
At large u, $T_{2\pi}(u)\approx1$ and (\ref{metric}) reduces to $AdS_{d+2}$ in Poincar\'e coordinates. However one of the spatial boundary coordinates $\theta$ is compactified on a circle. The remaining boundary coordinates $w_i$ and $t$ remain non-compact. The dual boundary theory may be thought of as a Scherk-Schwarz compactification on the $\theta$ cycle. The $\theta$ cycle shrinks to zero at a finite value of $u$, smoothly cutting off the IR region of $AdS$. This cutoff dynamically generates a mass scale in the theory, very much like in real QCD. The resulting theory is confining and has a mass gap.

Here we will work with\footnote{The analysis for any other $d \ge 3$ proceeds along the same lines and almost identical results can be obtained.} $d=4$ and make a coordinate transformation  $u=u_0+ax^2$ with $u_0=\log(2/5)$ and $a=5/4$, such that $T_{2\pi}(u_0)=0$ and for small $x$ the $x$-$\theta$ part of the metric looks flat, $ds^2\approx dx^2+x^2\,d\theta^2$. In these coordinates the metric is

\begin{align}
& ds^2=e^{2u_0+2ax^2}(-dt^2+T(x) d\theta^2+dw_i^2)+\frac{4a^2x^2}{T(x)} dx^2\,,\nn \\
{\rm where\ } & \ T(x)=1-e^{-5 a x^2}\,.
\label{metric1}
\end{align}

%

\subsection{Classical string in $AdS$-soliton}\label{subsec:clas}

We start with the Polyakov action: 
\beq
S_P = -\frac{1}{2\pi \al'}\int d\tau d\sigma \sqrt{-\ga} \ga^{ab}G_{\mu \nu}\dee_a X^\mu \dee_b X^\nu
\label{cstringS}
\eeq
where  $X^\mu$ are the coordinates of the string, $G_{\mu\nu}$ is the spacetime metric of the fixed background, $\gamma_{ab}$ is the worldsheet metric, the indices $a,b$ represent the coordinates on the worldsheet of the string which we denote as $(\tau,\sigma)$. We work in the conformal gauge $\gamma_{ab}=\eta_{ab}$ and use the following embedding for a closed string (partially motivated by \cite{Zayas:2010fs}):
\begin{align}
& t=t(\tau),\ \theta = \theta(\tau),\ x=x(\tau), \nn \\
& w_1=R(\tau)\cos\left(\phi(\sigma)\right),\ w_2=R(\tau)\sin\left(\phi(\sigma)\right)\ {\rm with\ } \phi(\sigma) = \alpha \sigma\,.
\label{embedding}
\end{align}
The string is at located at a certain value of $u$ and is wrapped around a pair of $w$-directions as a circle of radius $R$. It is allowed to move along the potential in $u$ direction and change its radius $R$. Here $\alpha\in{\mathbb Z}$ is the winding number of the string.
The test string Lagrangian takes the form:
\begin{align}
{\cal L}& \propto \frac{2}{5}e^{2ax^2}\left\{-\dot{t}^2+T(x)\dot{\theta}^2+\dot{w}_i^2-w_i'^2\right\} + \frac{2 a^2 x^2 }{T(x)} \dot x^2 \\
&=\frac{2}{5}e^{2ax^2}\left\{-\dot{t}^2+T(x)\dot{\theta}^2+\dot{R}^2-R^2\alpha^2\right\} + \frac{2 a^2 x^2 }{T(x)} \dot x^2\,,
\label{lag}
\end{align}
where dot and prime denote derivatives w.r.t $\tau$ and $\sigma$ respectively.  
The coordinates $t$ and $\theta$ are {\em ignorable} and the corresponding momenta are constants of motion. The test string Lagrangian differs from a test particle Lagrangian because of the potential term in $R(\tau)$. The coordinate $R$ would be ignorable without a potential term. In general it can be easily argued that for a generic motion of a test particle in an $AdS$ soliton background, all the coordinates other than $x$ are ignorable and the equations of motion can be reduced to a Lagrangian dynamics in one variable $x$. This implies integrability.

Here the conserved momenta conjugate to $t$ and $\theta$ are,
\begin{align}
p_t &= -\frac{4}{5}e^{2ax^2}\dot{t} \equiv-E \nn \\
p_{\theta} &= \frac{4}{5}e^{2ax^2}T(x)\dot{\theta} \equiv k. 
\label{Ek}
\end{align}
The conjugate momenta corresponding to the other coordinates are:
\begin{align}
p_R &=\frac{4}{5}e^{2ax^2}\dot{R} \nn \\
p_x &=\frac{4a^2x^2}{T(x)}\dot{x}\,.
\label{pxpR}
\end{align}
With these we can construct the Hamiltonian density:
\beq
\label{ham}
{\cal{H}} =\frac{5}{8} \left\{
\left(-E^2+\frac{k^2}{T(x)}+p_R^2\right)e^{-2ax^2}
+\frac{T(x)p_x^2}{5a^2x^2}
+\frac{16}{25}R^2\alpha^2e^{2ax^2}
\right\}
\eeq
Hamilton's equations of motion give:
\begin{align}
\dot{R} &= \frac{5}{4} p_R e^{-2ax^2} \\
\dot{p_R} &= -\frac{4}{5} R\alpha^2e^{2ax^2} \\
\dot{x} &= \frac{T(x)p_x}{4a^2x^2} \\
\dot{p_x} &= -\frac{5}{8} \left\{
4ax\left[\left(E^2-\frac{k^2}{T(x)}-p_R^2\right)e^{-2ax^2} + \frac{16}{25}R^2\alpha^2e^{2ax^2} \right] \right.\nn \\
&\left.-\frac{2T(x)p_x^2}{5a^2x^3} + \left[ \frac{p_x^2}{5a^2x^2}-\frac{k^2e^{-2ax^2}}{T(x)^2} \right]\partial_xT(x)
\right\} 
\label{eom}
\end{align}
We also have the constraint equations:
\begin{align}
\label{constraint}
& G_{\mu\nu}\left(\partial_\tau X^\mu\partial_\tau X^\nu+\partial_\sigma X^\mu\partial_\sigma X^\nu\right)=0\,, \\
& G_{\mu\nu}\partial_\tau X^\mu\partial_\sigma X^\nu=0\,.
\label{ceq}
\end{align}
The first equation takes the form ${\cal H}=0$ \footnote{The Hamiltonian constraint could be tuned to a nonzero value by adding a momentum in a decoupled compact direction. For example if the space is ${\cal M}\times S^5$ then giving a non-zero angular momentum in an $S^5$ direction would do the job. However we choose to confine the motion within $\cal M$ here.} and the second equation is automatically satisfied for our embedding.

\section{Dynamics of the system}\label{sec:dyn}
\begin{figure}[h!]
\centering
\subfigure[Periodic motion showing $R(\tau)\, ,x(\tau)$.]{
\includegraphics[scale=0.82]{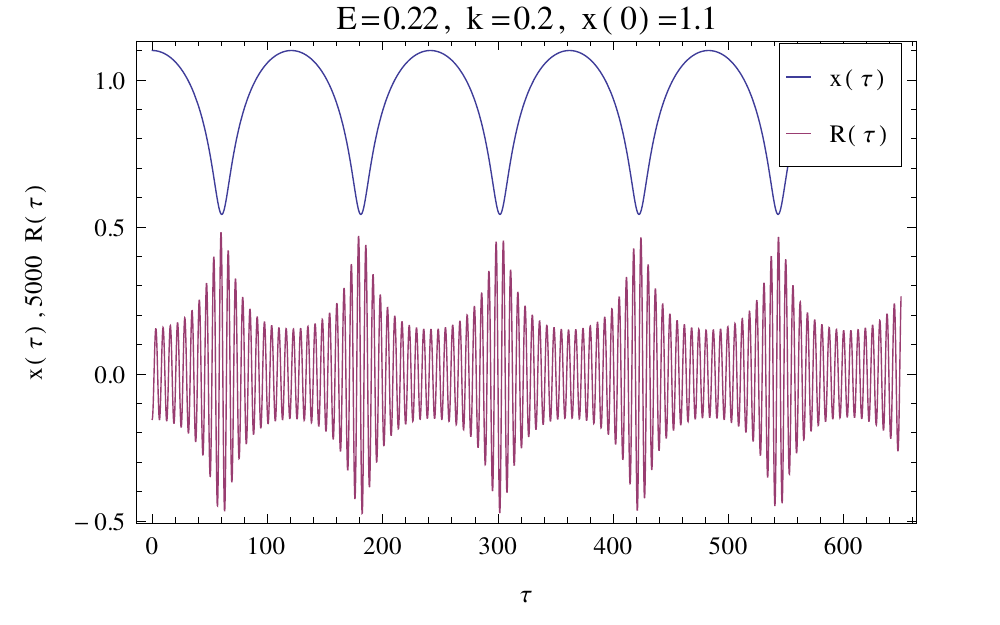}
\label{fig:mot_nc}
}
\subfigure[Power spectrum of $x(\tau)$ for periodic motion.]{
\includegraphics[scale=0.73]{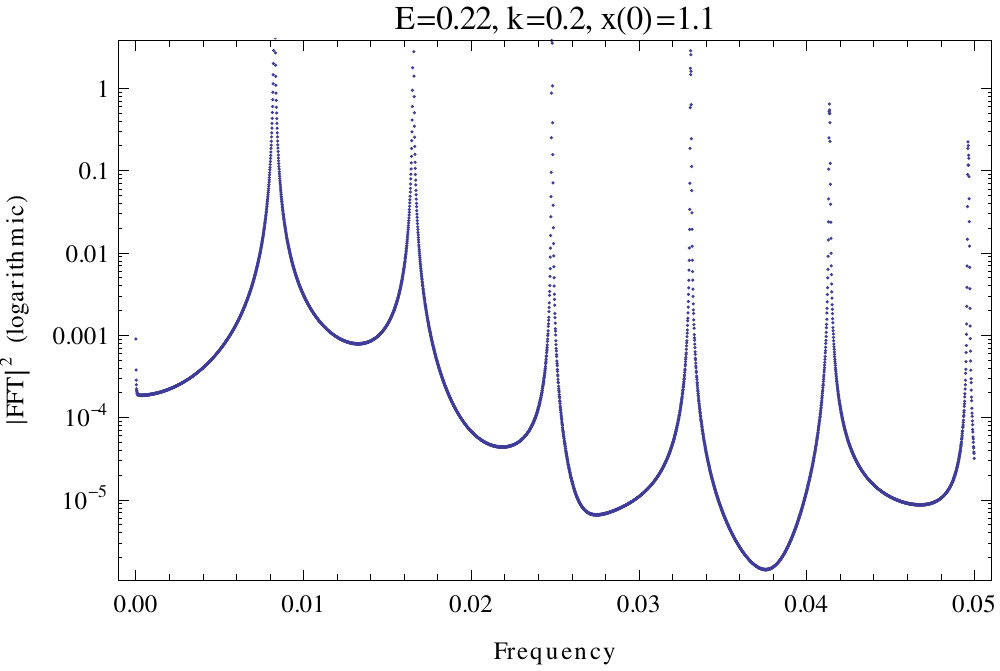}
\label{fig:fft_nc}
}
\subfigure[Chaotic motion of string showing $x(\tau)$.]{
\includegraphics[scale=0.75]{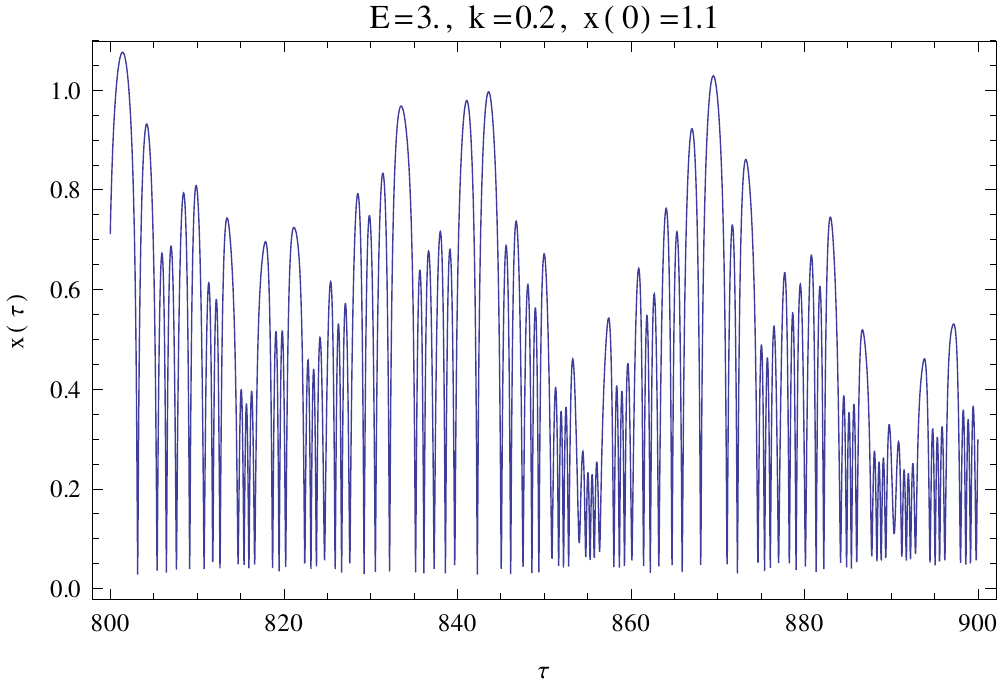}
\label{fig:mot_ch}
}
\subfigure[Power spectrum of $x(\tau)$ for chaotic motion.]{
\includegraphics[scale=0.75]{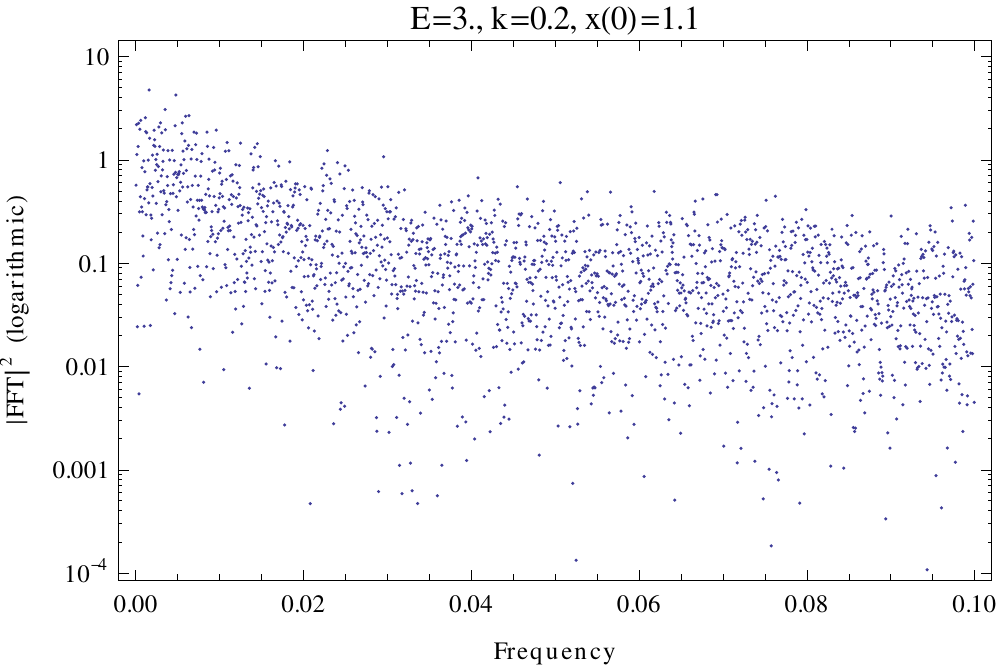}
\label{fig:fft_ch}
}
\caption{Numerical simulation of the motion of the string and the corresponding power spectra for small and large values of $E$. The initial momenta $p_x(0),p_R(0)$ have been set to zero. For a small value of $E=0.22$, we see a (quasi-)periodicity in the oscillations. The power spectrum shows peaks at discrete harmonic frequencies. However for a larger value of $E=3.0$, the motion is no longer periodic. We only show $x(\tau)$ but $R(\tau)$ is similar. The power spectrum is white.}
\label{fig:motion}
\end{figure}

At $k=0$, an exact solution to the EOM's is a fluctuating string at the tip of the geometry, given by
\begin{align}
x(\tau) &=0 \\
R(\tau) &=A \sin(\tau+\phi).
\end{align}
where $A,\phi$ are integration constants.  No such solution with constant $x(\tau)$ exists for $k \neq 0$. However one may construct approximate quasi-periodic solutions for small $R(\tau),p_R(\tau)$. It should be noted that with $R,p_R=0$ the zero energy condition Eqn.(\ref{constraint}) becomes similar to the condition for a massless particle and the string escapes from $AdS$ following a null geodesic. For small nonzero values of $R_0,p_R$, the motion in the $x$-direction will have a long time period. However the fluctuations in the radius will have a frequency proportional to the winding number which is of $\mo(1)$. This is a perfect setup to do a two scale analysis. In the equation for $\dot p_R$ we may replace $R(\tau)^2$ by a time average value. With this approximation, motion in the $x$-direction becomes an anharmonic problem in one variable which is solvable in principle. The motion is also periodic [Fig.\ref{fig:mot_nc}]. On the other hand to solve for $R(\tau)$ we treat $x(\tau)$ as a slowly varying field. In this approximation the solution for $R(\tau)$ is given by
\begin{align}
R(\tau) \approx \exp(-a\,x(\tau)^2) A \sin(\tau+\phi).
\end{align}
Hence $R(\tau)$ is quasi-periodic [Fig.\ref{fig:mot_nc}]. We have verified that in the small $R$ regime, the semi-analytic solution matches quite well with our numerics. 

Once we start moving away from the small $R$ limit the the above two scale analysis breaks down and the nonlinear coupling between two oscillators gradually becomes important. In short the coupling between oscillators tends to increase as we increase the energy of the string. Due to the nonlinearity, the fluctuations in the $x$- and $R$-coordinates influence each other and the motions in both coordinates become aperiodic. Eventually the system becomes completely chaotic [Fig.\ref{fig:mot_ch}]. The power spectrum changes from peaked to noisy as chaos sets in [Fig.\ref{fig:motion}]. As we discuss in the next subsection, the pattern follows general expectations from the KAM theorem. 

\subsection{Poincar\'e sections and the KAM theorem} \label{subsec:poin}

An integrable system has the same number of conserved quantities as degrees of freedom. 
A convenient way to understand these conserved charges is by looking at the phase space. Let us assume that we have a system with $N$ position variables $q_i$ with conjugate momenta $p_i$. The phase space is $2N$-dimensional. Integrability means that there are $N$ conserved charges $Q_i=f_i(p,q)$ which are constants of motion. One of them is the energy. These charges define a $N$-dimensional surface in the phase space which is a topological torus (KAM torus). The $2N$-dimensional phase space is nicely foliated by these $N$-dimensional tori. In terms of action-angle variables ($I_i,\theta_i$) these tori just become surfaces of constant action. With each torus there are $N$ associated frequencies $\omega_i(I_i)$, which are the frequencies of motion in each of the action-angle directions. 
 
It is interesting to study what happens to these tori when an integrable Hamiltonian is perturbed by a small nonintegrable piece. The KAM theorem states that most tori survive, but suffer a small deformation \cite{Ott,Hilborn}. However the {\em resonant} tori which have rational ratios of frequencies, i.e. $m_i\omega_i=0$ with $m\in {\cal Q}$, get destroyed and motion on them become chaotic. For small values of the nonintegrable perturbations, these chaotic regions span a very small portion of the phase space and are not readily noticeable in a numerical study. As the strength of the nonintegrable interaction increases, more tori gradually get destroyed. A nicely foliated picture of the phase space is no longer applicable and the trajectories freely explore the entire phase space with energy as the only constraint. In such cases the motion is completely chaotic.

To numerically investigate this gradual disappearance of foliation we look at the Poincar\'e sections. For our system, the phase space has four variables $x,R,p_x,p_R$. If we fix the energy we are in a three dimensional subspace. Now if we start with some initial condition and time-evolve, the motion is confined to a two dimensional torus for the integrable case. This 2d torus intersects the $R=0$ hyperplane at a circle. Taking repeated snapshots of the system as it crosses $R=0$ and plotting the value of $(x,p_x)$, we can reconstruct this circle. Furthermore varying the initial conditions (in particular we set $R(0)=0$, $p_x(0)=0$, vary $x(0)$ and determine $p_R(0)$ from the energy constraint), we can expect to get the foliation structure typical of an integrable system. 

\begin{figure}[h!]
\centering
\subfigure[$E = 0.25$]{
\includegraphics[scale=0.45]{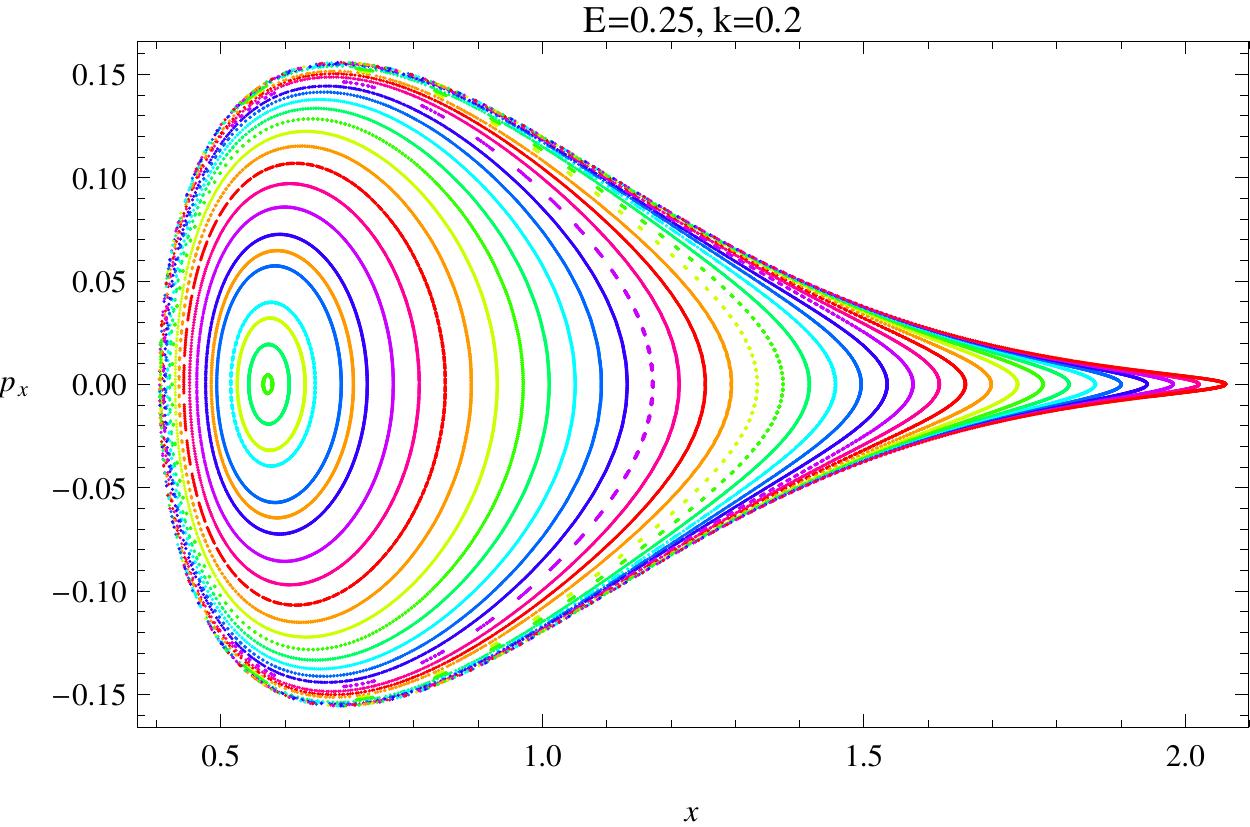}
\label{fig:psec1}
}
\subfigure[$E = 0.30$]{
\includegraphics[scale=0.45]{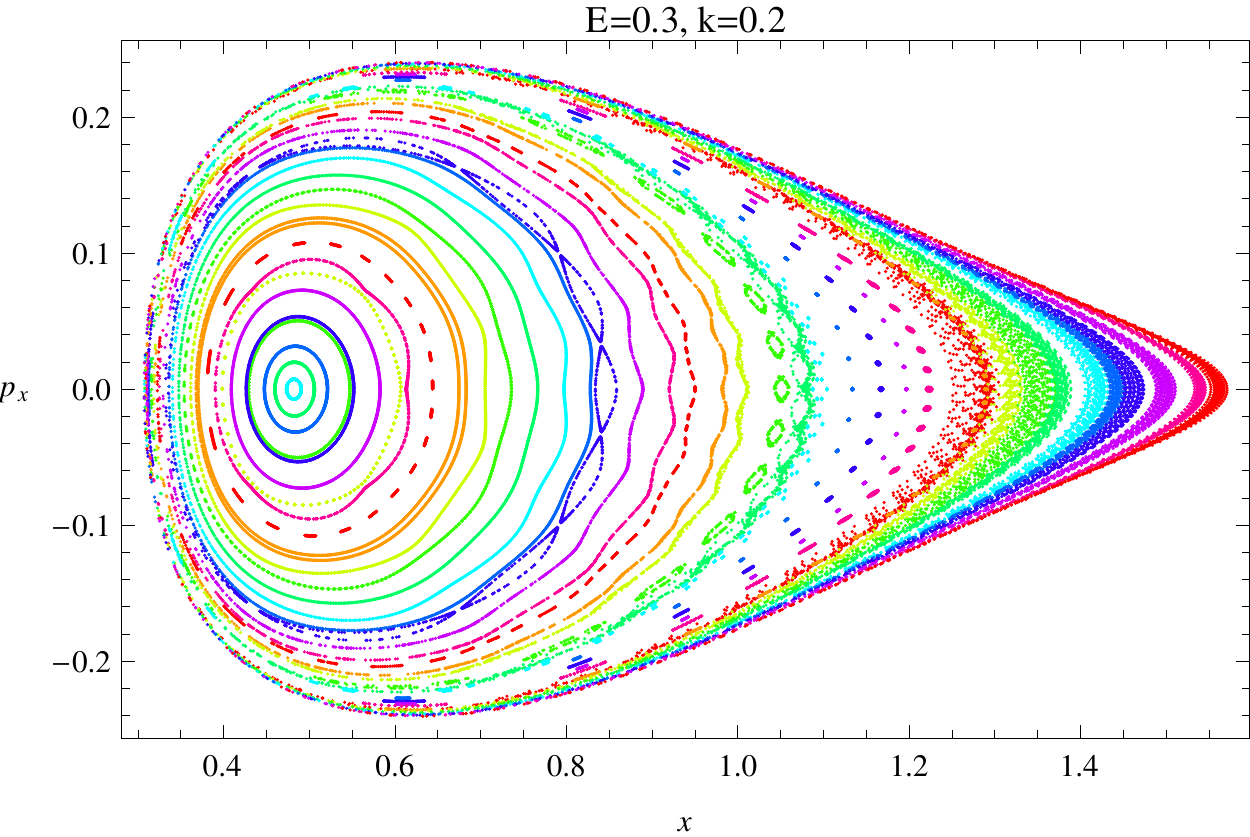}
\label{fig:psec2}}
\subfigure[$E = 0.36$]{
\includegraphics[scale=0.45]{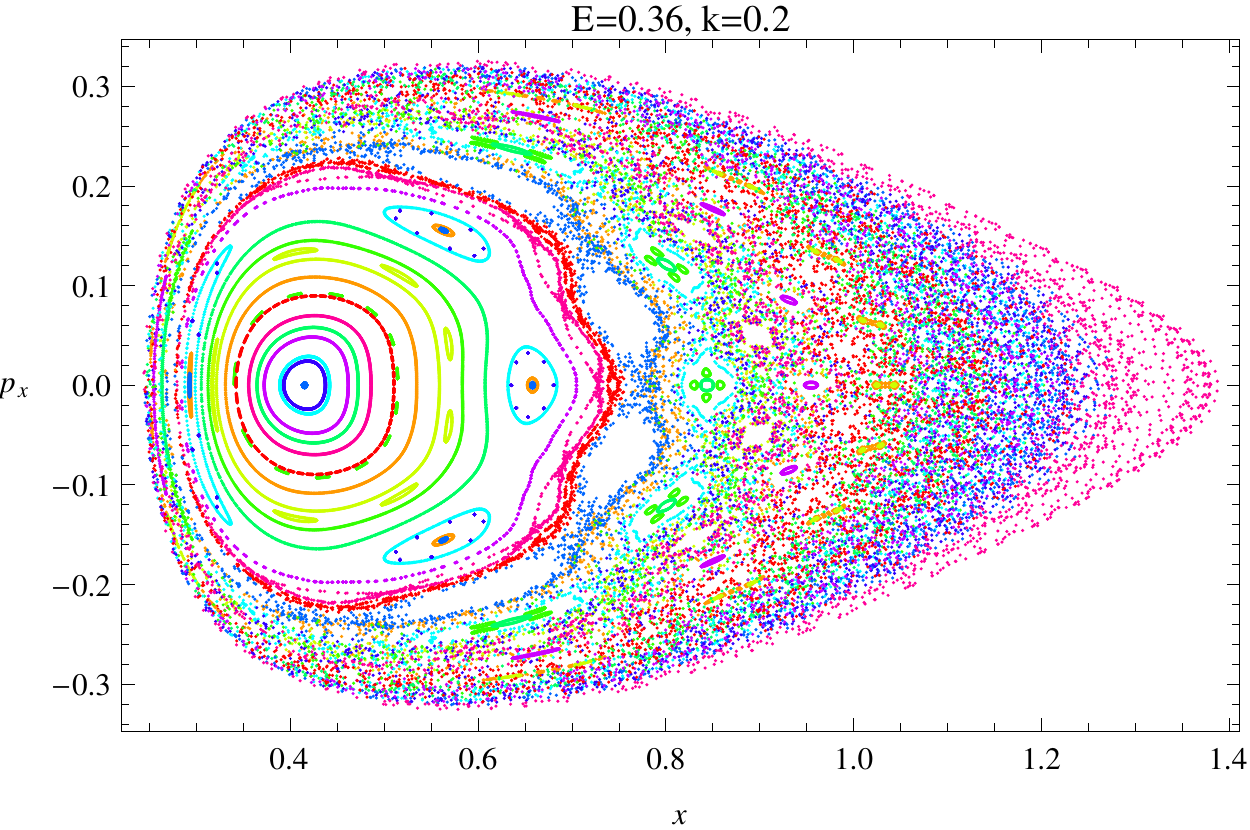}
\label{fig:psec3}
}
\subfigure[$E = 0.38$]{
\includegraphics[scale=0.45]{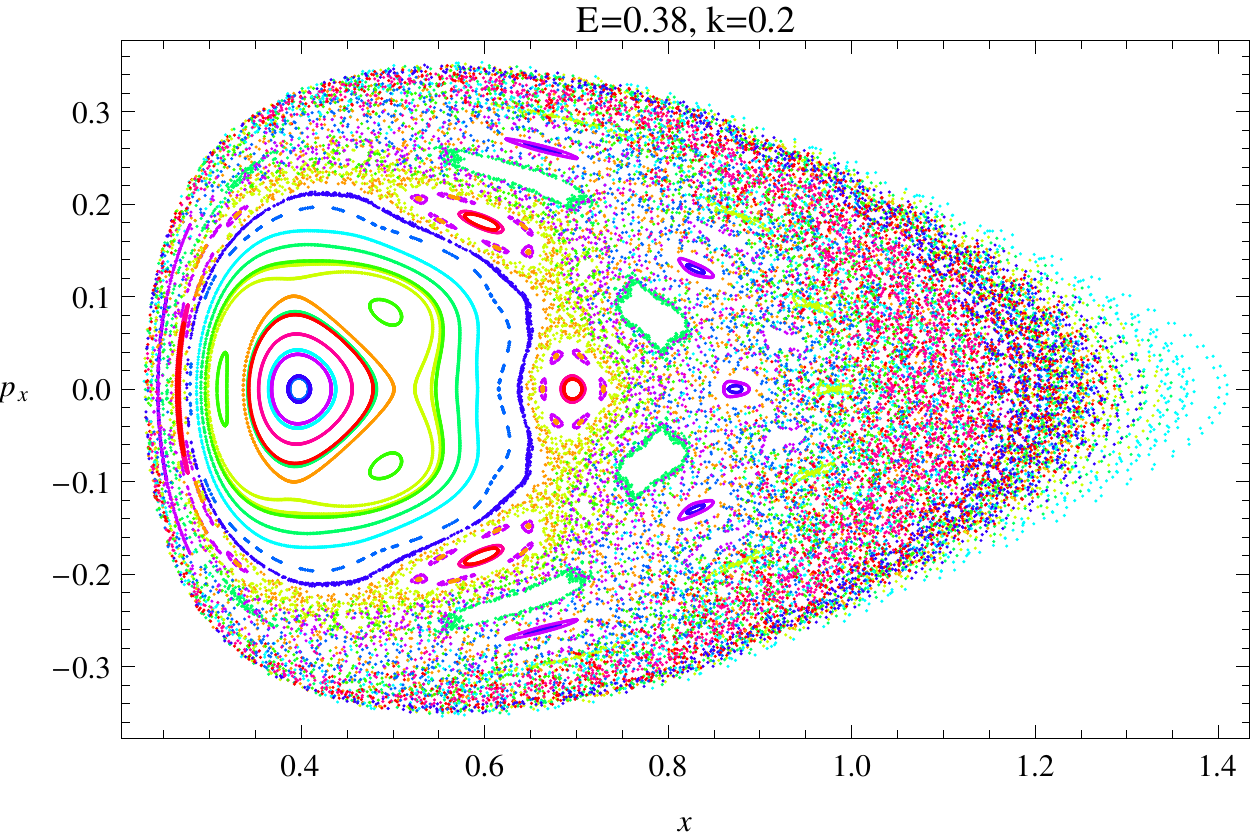}
\label{fig:psec4}
}
\subfigure[$E = 0.42$]{
\includegraphics[scale=0.45]{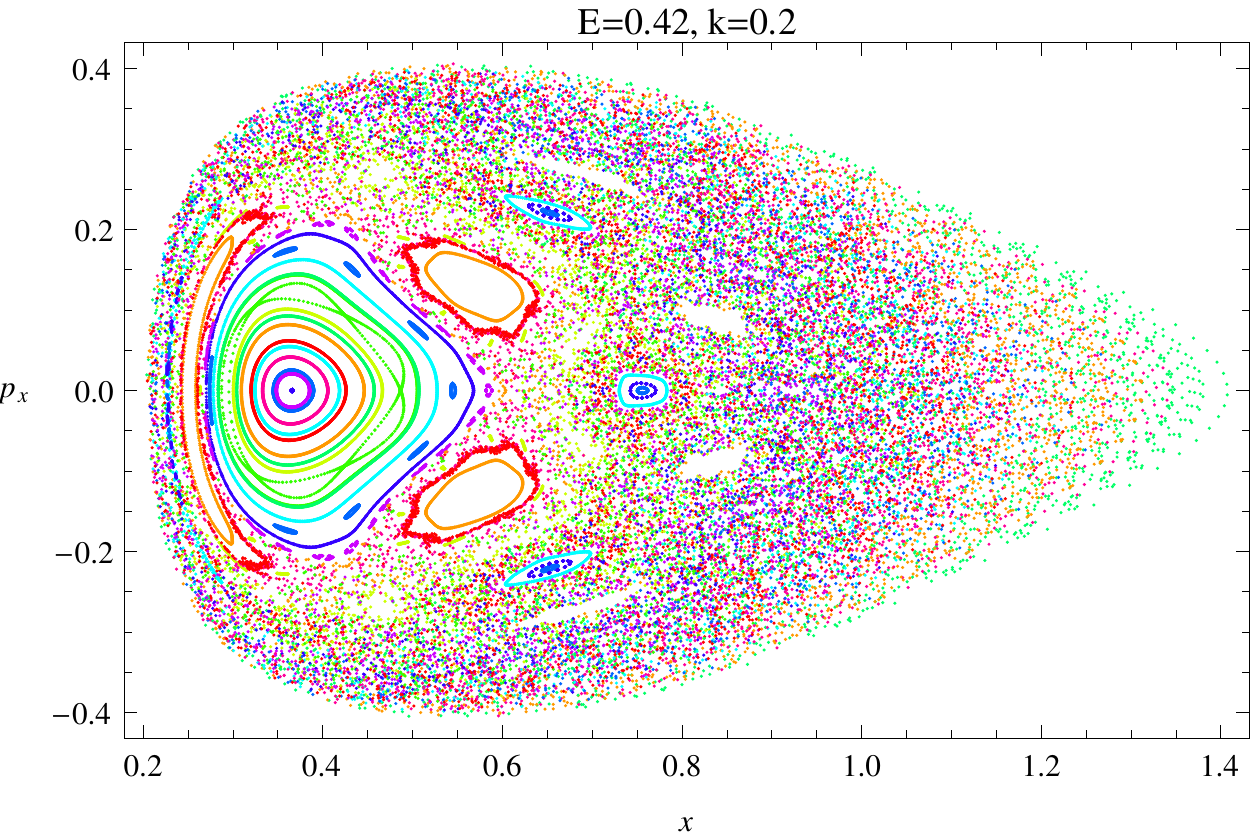}
\label{fig:psec5}
}
\subfigure[$E = 0.60$]{
\includegraphics[scale=0.45]{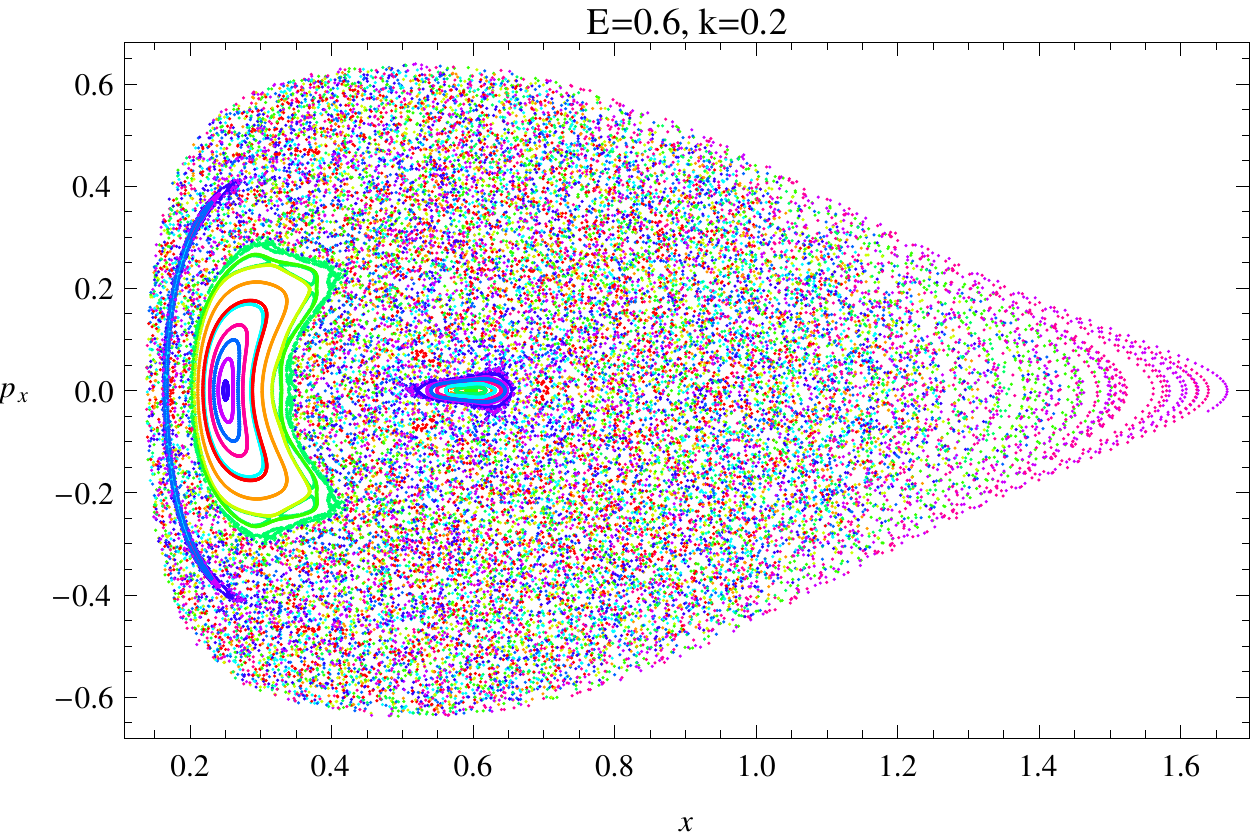}
\label{fig:psec6}
}
\subfigure[$E = 1.0$]{
\includegraphics[scale=0.45]{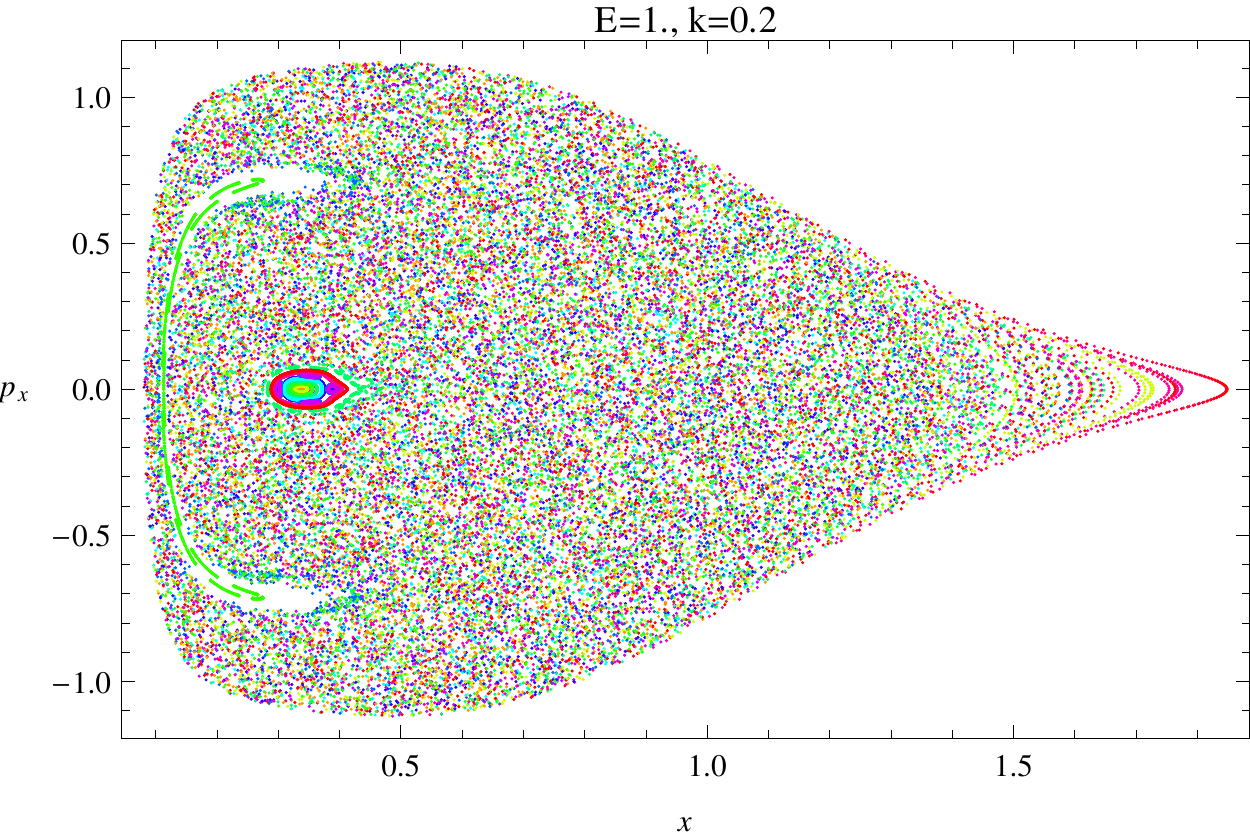}
\label{fig:psec7}
}
\subfigure[$E = 3.0$]{
\includegraphics[scale=0.45]{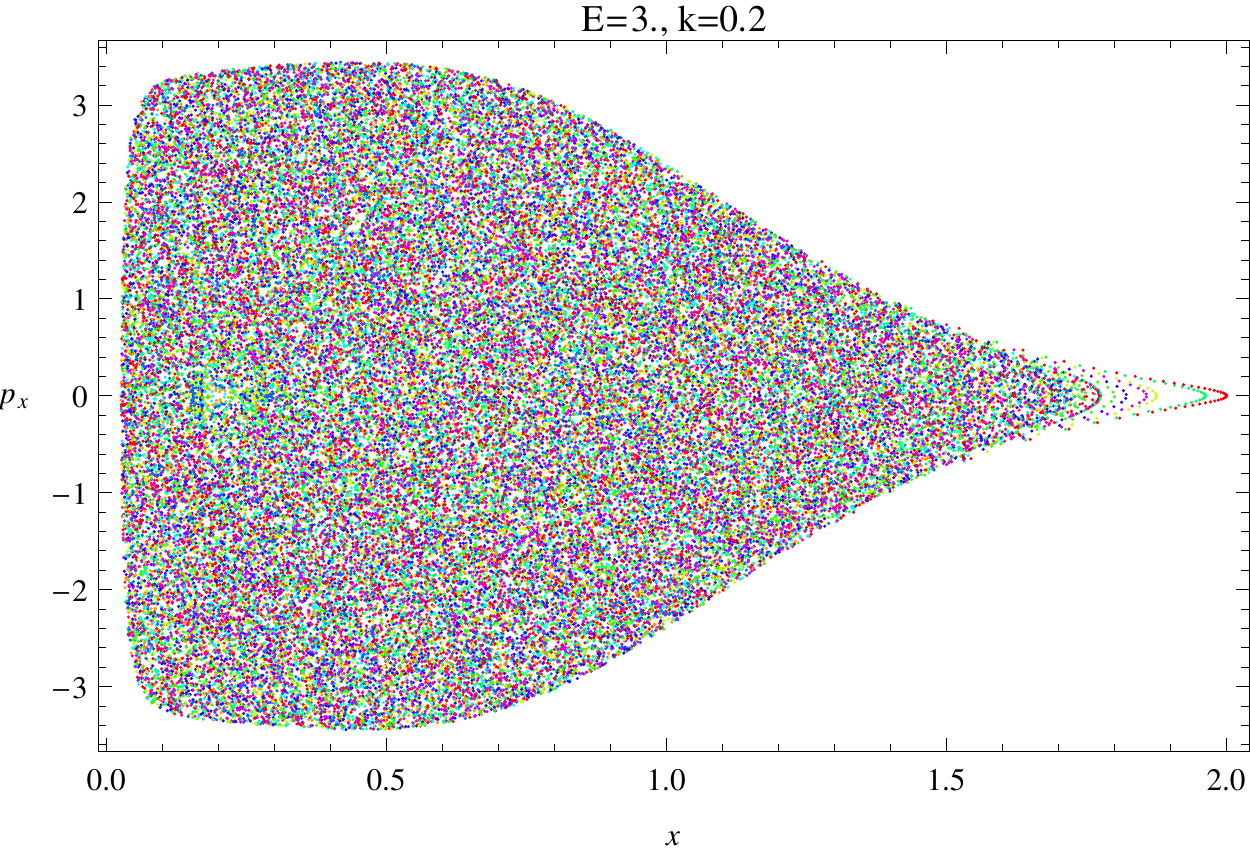}
\label{fig:psec8}
}
\caption{Poincar\'e sections demonstrate breaking of the KAM tori en route to chaos. Each colour represents a different initial condition. For smaller values of $E$ the sections of the KAM tori are intact curves, except for the resonant ones. The tori near the resonant ones start breaking as $E$ is increased. For very large values of $E$ all the colours get mixed -- this indicates that all the tori get broken and they fill the entire phase space.}

\end{figure}

Indeed we see that for smaller value of energies, a distinct foliation structure exists in the phase space [Fig.\ref{fig:psec1}]. However as we increase the energy some tori get gradually dissolved [Figs.\ref{fig:psec2}-\ref{fig:psec6}]. The tori which are destroyed sometimes get broken down into smaller tori [Figs.\ref{fig:psec3}-\ref{fig:psec4}]. Eventually the tori disappear and become a collection of scattered points known as cantori. However the breadths of these cantori are restricted by the undissolved tori and other dynamical elements. Usually they do not span the whole phase space [Figs.\ref{fig:psec3}-\ref{fig:psec6}].  For sufficiently large values of energy there are no well defined tori. In this case phase space trajectories are all jumbled up and trajectories with very different initial conditions come arbitrary close to each other [Fig.\ref{fig:psec8}].  The mechanism is very similar to what happens in well known nonintegrable systems like H\'enon-Heiles models \cite{Ott,Hilborn}.

\subsection{Lyapunov exponent} \label{subsec:lya}
\begin{figure}[h!]
\centering
\subfigure[Lyapunov index for periodic motion.]{
\includegraphics[scale=0.70]{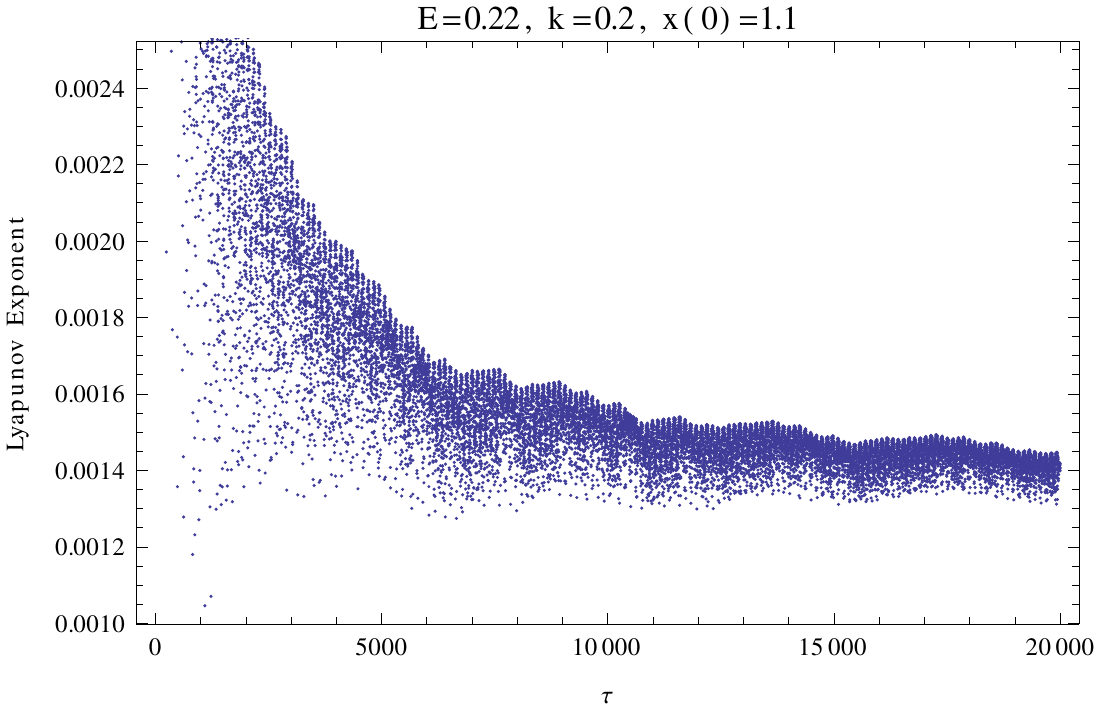}
\label{fig:le_nc}
}
\subfigure[Lyapunov index for chaotic motion.]{
\includegraphics[scale=0.60]{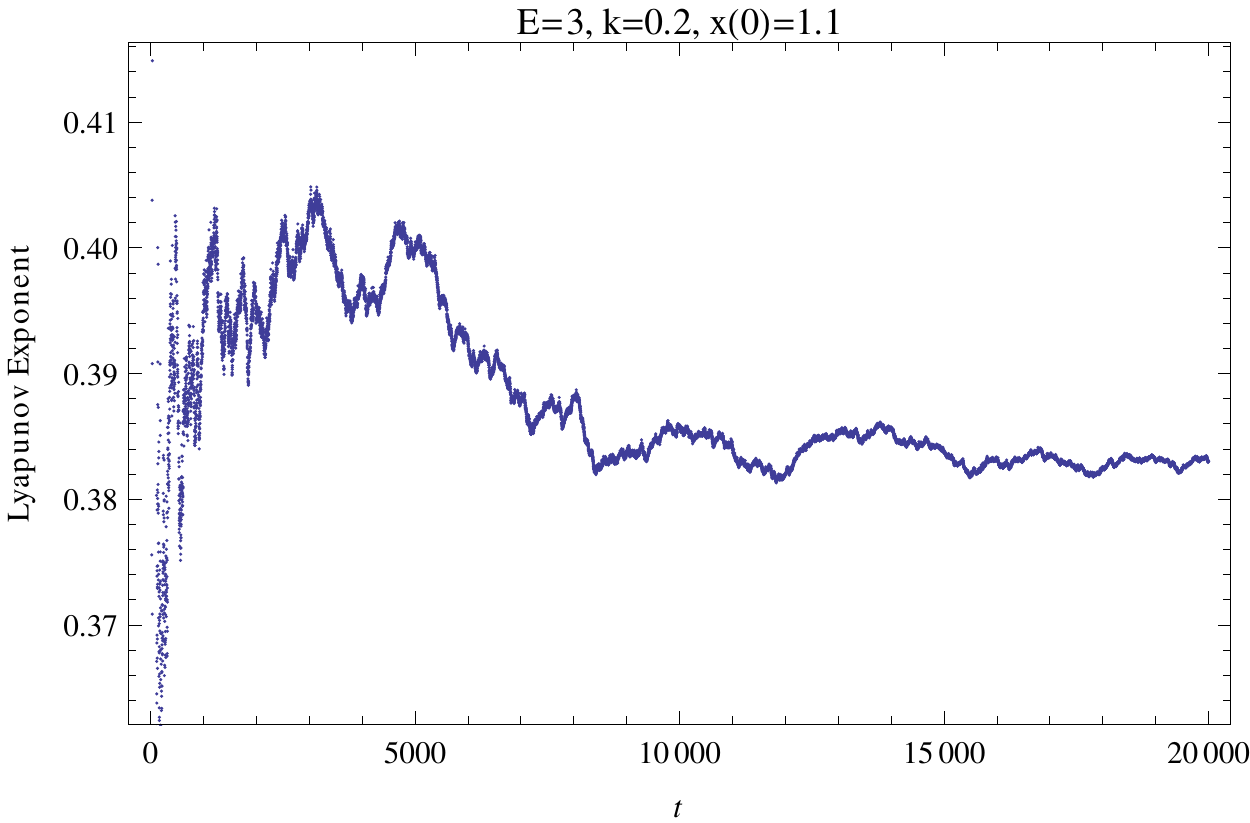}
\label{fig:le_ch}
}
\caption{Lyapunov indices for the same values of parameters as in Fig.(\ref{fig:motion}). For $E=0.22$, the Lyapunov exponent falls off to zero. For $E=3.0$, the Lyapunov exponent converges to a positive value of about $0.38$.}
\label{fig:le}
\end{figure}

One of the trademark signatures of chaos is the sensitive dependence on initial conditions, which
means that for any point $X$ in the phase space, there is (at least) one point arbitrarily close to $X$ that diverges from $X$. The separation between the two is also a function of the initial location  and has the
form $\Delta X(X_0,\tau)$. The Lyapunov exponent is a quantity that characterizes the rate of separation of such infinitesimally close trajectories. Formally it is defined as,
\begin{align}
\lambda=\lim_{\tau\rightarrow \infty} \lim_{\Delta X_0 \rightarrow 0} \frac{1}{\tau} \ln \frac{\Delta X(X_0,\tau)}{\Delta X(X_0,0)}
\label{lyapunov}
\end{align}
In practice we use an algorithm by Sprott \cite{Sprott}, which calculates $\lambda$ over short intervals and then takes a time average. We should expect to observe that, as time $\tau$ is increased, $\lambda$ settles down to oscillate around a given value. For trajectories belonging to the KAM tori, $\lambda$ is zero, whereas it is expected to be non-zero for a chaotic orbit. We verify such expectations for our case. We calculate $\lambda$ with various initial conditions and parameters. For apparently chaotic orbits we observe a nicely convergent positive $\lambda$ [Fig.\ref{fig:le}].

\section{Conclusion} \label{sec:con}
In this work, we argue using numerical techniques that classical string motion in the $AdS$ soliton background is nonintegrable. This certainly restricts the solvability of such theories. Also our results give a perspective on how much of classical integrability may be extended to various holographic backgrounds, especially those with less symmetry than $AdS_5$. Non-integrability is possibly quite generic and might be demonstrated by studying time evolution of simple string configurations. In particular, the basic construction of our work seems to be extendible to other confining backgrounds \cite{Klebanov:2000hb}. It would be nice to explore these directions. Another interesting extension would be to include world sheet fermions. It is also to be kept in mind that the $AdS$ soliton is not an exact string background and possibly has $\alpha'$ corrections. However these effects are unlikely to change the main result of the current work. 

One big question is the implication of our result for the full quantum spectrum. This is in turn connected with the glueball spectrum of the dual theory. The low-lying string modes will possibly be decoupled from the center of mass motion and will be more like the flat space counterpart. However the higher modes will be affected by the nonlinearity. Many exact results are known for the quantum spectrum of a chaotic theory. It would be interesting to explore how these results apply in a mini-superspace quantization of our system.
 
It would also be interesting understand more on the gauge theory side \cite{Berenstein:2004ys}. The dual gauge theory is the ${\cal N}=4$ SYM theory with one compact direction with aperiodic boundary condition for fermions. This breaks supersymmetry and the low energy dynamics of the theory is confining. Here, a simple change in boundary condition is changing the integrability of the theory. It is also to be noted that the full SYM theory with $\frac{1}{N}$ corrections is almost surely nonintegrable.  Any apparent integrability would then be a property of the large-$N$ saddle points. 

\subsection*{Acknowledgements}

We thank Sumit Das, Justin David, D. Reichmann, Spenta Wadia, Leopoldo Pando Zayas and in particular Al Shapere for valuable comments and discussions. PB thanks the MCTP, Ann Arbor  and TIFR, Mumbai for hospitality where some part of this work was done. PB and AG are partially supported by grant NSF-PHY-0855614 and NSF-PHY-0970069.

\bibliographystyle{unsrt}
\bibliography{chaos}
\end{document}